\documentstyle[tighten,preprint,aps]{revtex}

\begin{document}
\draft
\newcommand{\be}{\begin{equation}}
\newcommand{\ee}{\end{equation}}
\newcommand{\bea}{\begin{eqnarray}}
\newcommand{\eea}{\end{eqnarray}}
\newcommand{\ddl}[1]{\stackrel{\leftarrow}{\partial\over\partial#1}}
\newcommand{\ddr}[1]{\stackrel{\rightarrow}{\partial\over\partial #1}}
\newcommand{\ddd}[2]{{\partial^2\over\partial #1\partial #2}}

\preprint{Nonlinearity {\bf 9}, 1023-47 (1996)}

\title
{Edge Diffraction, Trace Formulae and the Cardioid Billiard}

\author{Henrik Bruus}
\address{Centre de Recherches sur les Tr\`es Basses Temp\'eratures,
CNRS, BP 166, F-38042 Grenoble C\'edex 9, France}

\author{Niall D. Whelan\cite{newadd}}
\address{Centre for Chaos and Turbulence Studies,
Niels Bohr Institute,
Blegdamsvej 17, DK-2100, Copenhagen \O, Denmark}
\date{February 19, 1996}
\maketitle

\begin{abstract}
We study the effect of edge diffraction on the semiclassical
analysis of two dimensional
quantum systems by deriving a trace formula 
which incorporates paths hitting any number
of vertices embedded in an arbitrary potential.  
This formula is used to study the cardioid billiard, which has 
a single vertex. The formula works well for most of the short 
orbits we analyzed but fails for a few diffractive orbits due 
to a breakdown in the formalism for certain geometries.
We extend the symbolic dynamics to account for diffractive orbits
and use it to show that in the presence of parity symmetry the
trace formula decomposes in an elegant manner such that 
for the cardioid billiard the
diffractive orbits have no effect on the odd spectrum.
Including diffractive orbits helps resolve peaks in the density of even
states but does not appear to affect their positions.
An analysis of the level statistics shows no
significant difference between spectra with and without diffraction.
\end{abstract}



\section{Introduction}

Periodic orbit theory \cite{gutz} provides a method of relating local,
canonically invariant information 
about classical periodic orbits to global quantum information
such as the density of states. However, this theory
must be extended if the classical mechanics is not defined due to 
discontinuities.
There is one class of discontinuity which is relatively mild in that all
trajectories are well defined but just their behavior changes abruptly at
some points in phase space.  Examples of this include grazing angles in
billiards \cite{creep,uzy} and the 
circle-straight joint in the Bunimovich stadium
\cite{ssclgr,alonso}.  A more
severe discontinuity is one in which some trajectories are undefined. Examples
of this include the vertex of a wedge \cite{vwr,ndw1,ps,ndw2}
and three body collisions \cite{tanner} since in 
neither case can we continue a trajectory through the discontinuity.
Other examples of discontinuities include scattering singularities \cite{seba},
flux tubes \cite{many,br} and small
scattering disks \cite{br,rww}. In each case, periodic orbit theory can be
extended by incorporation of so-called diffractive effects -- in the case of
vertices this is called edge diffraction.
To incorporate the effect of a discontinuity, one compares to the solution 
of the local scattering problem. For the wedge this was solved by
Sommerfeld \cite{somm} and the solution discussed in Refs.\cite{kel,Pauli}.

The structure of the paper is as follows. In Section II we derive a 
trace formula for diffractive orbits analogous to the Gutzwiller trace
formula for ordinary orbits. The amplitude of each diffractive orbit
is affected by the curvatures it experiences on the geometric part of
its path as well as by the diffraction at the vertex. As an
example, we employ the theory in a numerical study of the cardioid 
billiard \cite{rob} which is ergodic \cite{mark}, and in Section III
we discuss various properties of this billiard.  
The comparison of the theory and numerics takes place in Section IV. 
A brief analysis of the spectral statistics of the cardiod billiard is
presented in Section V, while Section VI contains the conclusion.

\section{Trace Formula for Diffractive Orbits}

The semiclassical formula for the trace $g(E)$ of the Green function
$G(E)$ of a chaotic Hamiltonian is \cite{gutz}
\be \label{tf_geom}
g(E) \equiv {\mbox{Tr}} G(E) = {1 \over i\hbar} \sum_\gamma 
{T_\gamma \over \sqrt{\Lambda_\gamma} \mp 1/\sqrt{\Lambda_\gamma}}
\exp\{i(S_\gamma/\hbar-\sigma_\gamma\pi/2)\},
\ee
where the sum is over all periodic orbits $\gamma$. The factors 
$T_\gamma$, $S_\gamma$, $\Lambda_\gamma$ and $\sigma_\gamma$ are the
canonically invariant periods, actions, stabilities and Maslov indices
of the orbit $\gamma$. (The canonical invariance of $\sigma_\gamma$ is
proved in Ref.~\cite{stevie}.) The
$-/+$ factor refers to whether the orbit is direct hyperbolic or inverse
hyperbolic.
We extract the density of states $\rho(E)$ from the trace
through the identity $\rho(E)=-\mbox{Im}[g(E)]/\pi$.
Periodic orbits are singled out because they have a stationary phase with 
respect to small deviations. The requirement of stationary phase can also
select other phase space structures \cite{balblok} and
we will henceforth refer to the orbits which enter
Eq.~(\ref{tf_geom}) as ``geometric orbits'' to distinguish from other
possibilities. 

Another process which can lead to a stationary phase is a trajectory which hits
a vertex. While not a classical trajectory, it is still a path in the sense
of path integrals \cite{feynhib} with an amplitude that can be found 
by comparison with the scattering solution of a wedge; 
we call such a path ``diffractive''.
The asymptotic (in $\hbar$) contribution to the Green function of the 
Scr\"odinger equation arising from
the path from $x'$ to $x$ via the vertex at $\xi$ is \cite{vwr,kel,Pauli}
\be \label{g_diff}
G_d(x,x',E) \approx {\hbar^2 \over 2m}
d(\theta,\theta')G_f(x,\xi,E)G_f(\xi,x',E),
\ee
where $G_f$ is the Green function in the absence of the wedge and 
$d(\theta,\theta')$ is called a diffraction constant. It equals
\be \label{dcons}
d(\theta,\theta') = -2{\sin(\pi/\nu) \over \nu}\left\{
{1 \over \cos(\pi/\nu) - \cos((\theta-\theta')/\nu)} \pm
{1 \over \cos(\pi/\nu) + \cos((\theta+\theta')/\nu)} \right\},
\ee
where $\theta$ and $\theta'$ are measured with respect to the wedge normal
(unlike in Refs.~\cite{ps,ndw2}) and $\nu=\alpha/\pi$, $\alpha$ being
the opening angle of the wedge as shown 
in Fig.~\ref{wedge}a. The $+/-$ sign refers to Neumann/Dirichlet
boundary conditions. 
We will mainly be interested in $\alpha=2\pi$ for which
\be \label{dspec}
d(\theta,\theta') = \sec\left({\theta-\theta' \over 2}\right) \mp
                    \sec\left({\theta+\theta' \over 2}\right).
\ee
Note that $d(-\theta,\theta')=\mp d(\theta,\theta')$
and that $d(0,0)$ equals 2 for Dirichlet boundary conditions and 
0 for Neumann. The factor $\hbar^2/2m$ appears in Eq.~(\ref{g_diff})
because we are using energy dependent Green functions; it is absent if we
use the Green functions of the Helmholtz equation in two dimensions. 
The Green functions of Eq.~(\ref{g_diff}) all have units 
$1/[Energy][Length]^2$.

For free motion in two dimensions, 
$G_f(x_2,x_1,E) = -i\hbar^2 H_0^{(+)}(k|x_2-x_1|)/8m$, where 
$H_0^{(+)}$ is the outgoing Hankel function and
$k=\sqrt{2mE}/\hbar$. This form of $G_f$ is assumed in the derivation of the 
diffraction constant. However Eq.~(\ref{g_diff}) is more general. 
In the presence of a potential with a vertex, we
do the local scattering problem by assuming the potential does not change much
in a typical wavelength. We would then 
call the directions of the trajectory when it enters and leaves the vertex
$\theta'$ and $\theta$ and use these in determining the diffraction constant
$d(\theta,\theta')$. Away from the vertex, we connect the outgoing free space 
Green functions to the relevant semiclassical ones for that 
potential. For example, Eq.~(\ref{g_diff}) is valid
for the problem of motion bounded within a wedge in the presence of gravity
\cite{wdglit} for which $G_f$ is more complicated. For that reason, we do
not assume billiard conditions in the subsequent discussion, although we
do restrict the discussion to two spatial dimensions.
The content of Eqs.~(\ref{g_diff}) and (\ref{dcons}) is that an orbit entering
the vertex can be continued out at any angle with a quantum amplitude given by
$d(\theta,\theta')$.  The contribution of such a diffractive orbit is of
order $\sqrt{\hbar}$ relative to that of a geometric orbit.
We now analyze this semiclassically to derive a trace formula in
analogy to Eq.~(\ref{tf_geom}).

If there are $n$ diffractions as sketched in Fig.~\ref{wedge}b,
Eq.~(\ref{g_diff}) generalises to 
\be \label{manydiff}
G_d(x,x',E) \approx\left\{\prod_i^n {\hbar^2 \over 2m}d_i\right\} 
G_f(x,\xi_n)G_f(\xi_n,\xi_{n-1})\cdots G_f(\xi_2,\xi_1)G_f(\xi_1,x')
\ee
where we have suppressed the energy dependence in $G_f$ and the $\theta$
dependence in the diffraction constants.
To obtain the trace, we first identify the points $x$ and $x'$ and then invoke
stationary
phase to require that the momenta also match smoothly - as for geometric
orbits.  Orbits which satisfy these constraints we call diffractive periodic
orbits. They can be found, in principle, by firing out trajectories at all 
angles from all the vertices and determining which ones return to a vertex. 
We impose no constraint on the momenta at the vertices and so allow any 
incoming or outgoing angles.

To proceed, we define local coordinates along the various classical
paths. At each point $x$, we take $z$ and $y$ to be the local coordinates
parallel and transverse to the path, respectively.
At each vertex we define local coordinates $\zeta_i$ and $\eta_i$ on the
the incoming path and $\zeta'_i$ and $\eta'_i$ on the outgoing path. 
The local transverse momentum at $x$ is $p$ and the local transverse 
momenta at the vertex $\xi_i$ are $\pi_i$ and $\pi'_i$. 
The constant energy approximation to the classical Green 
function from any $x_1$ to any $x_2$ is \cite{gutz}
\be \label{gf}
G_f(x_2,x_1,E) \approx {1 \over i\hbar}{1 \over \sqrt{2\pi i\hbar}}
D(x_2,x_1)\exp\{i(S/\hbar-\mu\pi/2)\}.
\ee
The action $S$ is evaluated along the path
and $\mu$ counts the caustics along the orbit. 
The factor $D$ equals \cite{gutz}
\bea
D(x_2,x_1) & = & {1 \over |\dot{z_2}\dot{z_1}|^{1/2}}
\left| -{\partial^2 S \over \partial y_2\partial y_1} \right|^{1/2}
\nonumber \\
          & = & {1 \over |\dot{z_2}\dot{z_1}|^{1/2}} 
\left| \left({\partial y_2 \over \partial p_1} \right)_{y_1}\right|^{-1/2},
\label{dfact}
\eea
where the subscript on the bracket of the second equation indicates that
we take the derivative of the final position with respect to the initial
momentum while holding the initial position fixed. This then defines a fan
of initial conditions radiating from the source point $x_1$.
This is the contribution of a single classical trajectory -- if there is more 
than one, we must sum over them.

Evaluation of the trace involves integrating along the periodic
orbit and transverse to it. The integral along the orbit can be done one arc
at a time and below we consider just the arc between $\xi_1$ and $\xi_2$.
Eqs.~(\ref{gf}) and (\ref{dfact}) imply
\be \label{ggdiff}
G_f(\xi_2,x)G_f(x,\xi_1) 
\approx \left({1 \over i\hbar}{1 \over \sqrt{2\pi i\hbar}}
\right)^2 D(\xi_2,x)D(x,\xi_1)\exp\left\{i\left[S_{21}/\hbar - 
(\mu_2+\mu_2)\pi/2\right]\right\},
\ee
where $S_{21}=S_2+S_1$ is the action of the path between $\xi_1$
and $\xi_2$ via $x$. At each point $x$ along the orbit, we calculate 
the transverse $y$ integral. The only
$y$ dependence, to leading order in $\hbar$, is in the action which we 
approximate as 
\be \label{sapp}
S_{21}\approx S^0_{21}+{1\over 2} 
{\partial^2 S_{21} \over \partial y^2}y^2.
\ee
The partial derivative is taken while holding the initial and final
coordinates $\eta'_1$ and $\eta_2$ fixed at zero. $S^0_{21}$ is the
action evaluated at $y=0$ and is independent of the position $z$
along the orbit.
The stationary phase integral yields
\be \label{statfaz}
\int_{-\infty}^{\infty}dy\exp\left(iS_{21}/\hbar\right) = 
\sqrt{2\pi i\hbar}
\left|{\partial^2 S_{21} \over \partial y^2}\right|^{-1/2}
\exp\left(i(S^0_{21}/\hbar - \sigma\pi/2)\right), 
\ee
where $\sigma$ is zero if the second derivative in Eq.~(\ref{sapp}) is
positive and is unity if the second derivative is negative.

We now seek to manipulate the various partial derivatives which come
from the two amplitude factors $D$ and from the stationary phase integral
(\ref{statfaz}). These all come with a power of $-1/2$ and with absolute
value signs. For purposes of manipulation, we neglect those for the moment
so the combination we need to analyse is
\be \label{combo}
\left({\partial y \over \partial \pi'_1}\right)_{\eta'_1}
\left({\partial y \over \partial \pi_2}\right)_{\eta_2}
{\partial^2 (S_1+S_2) \over \partial y^2}
=
\left({\partial y \over \partial \pi'_1}\right)_{\eta'_1}
\left({\partial y \over \partial \pi_2}\right)_{\eta_2}
\left(\left({\partial p \over \partial y}\right)_{\eta'_1}
-     \left({\partial p \over \partial y}\right)_{\eta_2}\right).
\ee
We have used the fact that the derivative of the action with respect
to the final position equals the final moment while the derivative with 
respect to the initial position equals the negative of the initial momentum.
Also, we have inserted all of the relevant subscripts to indicate what is
being kept fixed in each derivative. We now show that this factor is
independent of position along the orbit. We can combine partial
derivatives in Eq.~(\ref{combo}) to obtain
\be \label{comboprime}
\left({\partial y \over \partial \pi_2}\right)_{\eta_2}
\left({\partial p \over \partial \pi'_1}\right)_{\eta'_1}     -
\left({\partial y \over \partial \pi'_1}\right)_{\eta'_1}
\left({\partial p \over \partial \pi_2}\right)_{\eta_2}
=
\left({\partial(y,p) \over \partial(\pi_2,\pi'_1)}\right)_{\eta'_1,\eta_2}.
\ee
In the right half of Eq.~(\ref{comboprime}) we have borrowed the Jacobian
notation of Ref.~\cite{stevlit}. (We have made use of the trivial
freedom to specify that in the first derivative of the left hand side we are
also holding $\eta'_1$ fixed, with similar specifications in all four terms.)
To determine the value of this Jacobian
corresponding to some different point $z'$ along the trajectory, we should
multiply Eq.~(\ref{comboprime}) by the Jacobian relating the
transverse variables variables $(y,p)$ at position $z$ 
to the transverse variables $(y',p')$ position $z'$.
However, because these variables are transverse to a trajectory, their
Jacobian is identically unity; the two sets of
variables are canonically related owing to symplectic nature of the
Hamiltonian flow. It follows that the combination of factors 
appearing in Eq.(\ref{combo}) is independent of position $z'$.
In particular, it is particularly convenient to calculate it very close to
one of the vertices. If $z$ is such that the point is close to $\xi_2$, we
have
\be \label{simpfacs}
\left({\partial y \over \partial \pi_2}\right)_{\eta_2} = 0 \;\;\;\;\;\;\;\;
\left({\partial p \over \partial \pi_2}\right)_{\eta_2} = 1.
\ee
It follows that
\be \label{gettingthere}
\left({\partial(y,p) \over \partial(\pi_2,\pi'_1)}\right)_{\eta'_1,\eta_2} =
-\left({\partial y \over \partial \pi'_1}\right)_{\eta'_1},
\ee
where the right hand side is evaluated at the point $\xi_2$. Heceforth, we 
change notation slightly and call this term
${\partial\eta_2 \over \partial \pi'_1}$ to stress that it is evaluated at
the second vertex. This factor is simply the spread in position at $\xi_2$ 
of a fan of trajectories radiating from $\xi_1$.

We also want that the phase index $\mu_1+\mu_2+\sigma$ be independent of
position along the orbit. It is not true that the indices are separately
invariant; it is simple to imagine that as we change position $z$ along the
orbit, we will gain or lose caustics in going from the two vertices to the 
intermediate position. However, these changes will be exactly mirrored by
changes in the index $\sigma$ such that the sum is invariant. For a 
demonstration of this, we refer to Ref.\cite{grpar} where the authors
evaluate an integral similar to Eq.~(\ref{manydiff}). They interpret 
Eq.~(\ref{gf}) as a propagator along the orbit with $z$ playing the role
of time. Using the semigroup property of time-dependent propagators, they
conclude that the phase index is a constant. Since this is constant, we are 
free to use $\mu_{21}$ which is the number of caustics of a fan of trajectories
going from vertex 1 to vertex 2.

That completes the $y$ integral. For the $z$ integral, we remark that the 
only $z$ dependence is in the velocity $|\dot{z}|^{-1/2}$ which appears in
the amplitudes $D$ of Eq.~(\ref{ggdiff}). Since there are two of them, the 
integral to be performed is simply
\be \label{zint}
\int {dz \over |\dot{z}|} = T_{21}
\ee
which is just the time it takes to get from $\xi_1$ to $\xi_2$.
Putting together all the remaining factors, we conclude that
\bea 
\int dzdy G_f(\xi_2,x)G_f(x,\xi_1) & \approx & 
{T_{21} \over (i\hbar)^2\sqrt{2\pi i\hbar}}
{1 \over |\dot{\zeta}_2\dot{\zeta}_1|^{1/2}}
\left|{\partial\eta_2 \over \partial \pi'_1}\right|^{-1/2}
\exp\left\{i(S_{21}-\mu_{21}\pi/2)\right\} \nonumber\\
& = & {T_{21} \over i\hbar}G_f(\xi_2,\xi_1), \label{concl}
\eea
where we have used
Eqs.~(\ref{gf}) and (\ref{dfact}) in the second line and dropped the
superscript $0$ on the action.

The appealing fact that the trace integral on the arc between $\xi_1$ and 
$\xi_2$ is proportional to the Green function between these points
simplifies the analysis tremendously. Recall that we must still multiply all
of the other free Green functions from Eq.~(\ref{manydiff}) so there is
a factor which is simply the product of all the Green functions from vertex
to vertex.  In doing the integrals along the arc between $\xi_i$ and 
$\xi_{i+1}$, we get exactly the same product but multiplied by $T_{i+1,i}$ so
that the integral of Eq.~(\ref{manydiff}) is
\bea
g_{\gamma}(E) & = & {T_{\gamma} \over i\hbar}
\left\{\prod_{i=1}^{n_{\gamma}}
\left({\hbar^2 \over 2m}\right)d_i G(\xi_{i+1},\xi_i)\right\}
\nonumber\\
& = & {T_{\gamma} \over i\hbar} 
\left({\hbar\over 8\pi m^2}\right)^{n_{\gamma}/2}
\left\{\prod_{i=1}^{n_{\gamma}} {d_i \over |\dot{\zeta_i}|} 
\left|{\partial\eta_{i+1} \over \partial \pi'_i}\right|^{-1/2}\right\}
\exp\left\{i(S_{\gamma}/\hbar - 
\sigma_{\gamma}\pi/2 - 3n_{\gamma}\pi/4)\right\}, \label{tr1}
\eea
where $S_{\gamma}$, $T_{\gamma}$ and $\sigma_{\gamma}$ 
are the sums of $S_{i+1,i}$, $T_{i+1,i}$ and 
$\mu_{i+1,i}$ along the orbit. The velocity  $\dot{\zeta_i}$ is given by energy
conservation and is a constant at each vertex and the index $i$
is cyclic so vertex $n+1$ is identified with vertex 1.
This diffractive trace formula is the main result of this section.
The formula was given in Ref.~\cite{vwr} by comparison with
creeping diffraction, where
Watson contour integration \cite{creep} can be used to show that the trace 
has the same structure. The diffractive trace formula is similar to
the Gutzwiller trace formula Eq.~(\ref{tf_geom}) but is suppressed
by a relative factor of $\hbar^{n/2}$ \cite{balblok}.
Eq.~(\ref{tr1}) only shows the contribution of a single
diffractive orbit; in practice we must sum over all such orbits and so
introduce the subscript $\gamma$ to distinguish them. 
If the orbit is a repeat of a shorter
primitive orbit, the factor of $T$ in Eq.~(\ref{tr1}) is the period of the
primitive orbit.

We now specialise to the potential-free case
so that $\hbar=m=1$, $S/\hbar=kL$, $T=L/k$, $E=k^2/2$, and
$|\dot{\zeta}_i|=k$. We further invoke 
the infinitesimal relation $\delta\pi'_i=k\delta\phi'_i$ where 
$\delta\phi'_{i}$ is the angular deviation from the
periodic orbit on leaving vertex $\xi_{i}$
so that
\be \label{ddd}
\left|{\partial\eta_{i+1} \over \partial \pi'_{i}}\right| = {1 \over k}
\left|{\partial\eta_{i+1} \over \partial \phi'_{i}}\right| 
\equiv {1 \over k}F_i.
\ee
The contribution to the density of states in $k$ is given by
$\rho_{\gamma}(k) = -k\mbox{Im}[g_{\gamma}(E)]/\pi$ so
\be \label{finally}
\rho_{\gamma}(k) = {L_{\gamma} \over \pi} 
\left\{\prod_{i=1}^{n_{\gamma}} {d_i \over \sqrt{8\pi kF_i}}\right\}
\cos\left(kL_{\gamma}-\sigma_{\gamma}\pi/2-3n_{\gamma}\pi/4\right),
\ee
to be contrasted with 
\be \label{rhog}
\rho_g(k) = {L_g \over \pi} 
\frac{1}{\sqrt{\Lambda_g} \mp 1/\sqrt{\Lambda_g}}
\cos\left(kL_g-\sigma_g\pi/2\right)
\ee
for geometric orbits.
In Eq.~(\ref{finally}) the factor $F_i$ 
has a simple interpretation; if we launch a narrow cone
of trajectories from vertex $\xi_i$ centered on the periodic orbit, 
$F_i$ gives the width of the cone when it arrives at $\xi_{i+1}$ \cite{ndw1}. 
This interpretation in terms of cones is in contrast with that of cylinders 
for geometric orbits.  Eq.~(\ref{finally}) was also obtained in Ref.\cite{ps}
for the special case of straight walls everywhere so that
$F_i = L_i$, the distance between the vertices.  We also mention that this
analysis applies equally well if two of the diffraction points $\xi_i$ are
at the same vertex. In particular, if there is only one vertex,
then the diffractive periodic orbits are those which leave the vertex and
return to it following a classical path.
Eqs.~(\ref{tr1}) and (\ref{finally}) are true regardless of whether the
classical motion is chaotic or not, although they are restricted to
isolated diffractive orbits.

There is a zeta function \cite{ruelle} corresponding
to Eq.~(\ref{tr1}) in analogy to that which exists for geometric orbits
\cite{vor}. We start by defining the diffractive weight for each diffractive
orbit
\be \label{diffweight}
t_\gamma = \left({\hbar\over 8\pi m^2}\right)^{n_\gamma/2}
\left\{\prod_i^{n_\gamma} {d_i \over |\dot{\zeta_i}|} 
\left|{\partial\eta_{i+1} \over \partial \pi'_i}\right|^{-1/2}\right\}_\gamma
\exp\left\{i(S_\gamma/\hbar - \sigma_\gamma\pi/2 - 3n_\gamma\pi/4)\right\},
\ee
where we have now introduced the subscript $\gamma$ to distinguish diffractive
orbits. The sum over diffractive orbits is then
\be \label{sodo}
g_d(E) =  
\sum_\gamma {T_\gamma \over i\hbar}\sum_{r=1}^\infty t_\gamma^r
=\sum_\gamma {T_\gamma \over i\hbar}{t_\gamma \over 1-t_\gamma} ,
\ee
where we have organised the sum into the primitive orbits and their repeats.
To leading order in $\hbar$,
\be \label{tder}
{d t_\gamma \over dE} \approx -{T_\gamma \over i\hbar}t_\gamma
\ee
so that
\be \label{logder}
g_d(E) = {d\over dE}\log\left(\prod_\gamma (1-t_\gamma)\right).
\ee
The quantity $\zeta_d^{-1}(E)=\prod_\gamma (1-t_\gamma)$ is the
diffractive zeta function to the power $-1$.  When multiplied by the
corresponding geometric zeta function \cite{creep} to the power $-1$ and
appropriately regulated \cite{symbprune,regulate}, 
the product equals the spectral determinant $\prod_n(E-E_n)$
so that its zeros give the quantum energy levels.  Due to the
regularisation, the zeros of the separate terms in $\zeta_d^{-1}(E)$
are not true zeros of the product. A formula analogous to 
Eq.~(\ref{logder}) also holds in the case of billiards \cite{vwr,ndw2}.

The function $\zeta_d^{-1}(E)$ involves only a single
product over periodic orbits. In contrast,
the zeta function for geometric orbits 
has an additional product over an integer index which
can be thought of as labeling
local eigenstates transverse to each orbit \cite{eigenmodes}. 
Near the orbit, these local eigenstates typically have the form 
$\psi_n(y)\sim y^n$ with $n\geq 0$. We conclude that 
diffractive orbits have only the $n=0$ local eigenstate.
Higher states do not exist because they would
have a node on the periodic orbit and
would not be affected by the diffraction. This was also noted in the
scattering geometries discussed in Refs.~\cite{ndw1,ndw2} where it 
caused there to be no lower families of quantum resonances.
This difference is intimately related to the fact that the Green
functions in Eq.~(\ref{manydiff}) are multiplicative 
\cite{creep} in contrast to the
behavior of Green functions for geometric orbits \cite{gutz}.

\section{The Cardioid Billiard}

In this section, we discuss various aspects of the cardioid billiard which
are relevant to us.  We briefly review its classical properties and
construct the symbolic dynamics for the geometric orbits and diffractive
orbits. The Maslov indices are given by a simple rule in terms of the
symbolical dynamics.  We then
discuss the role of symmetry in the quantum problem and how the geometric
and diffractive trace formulas conspire to give the even and odd spectra.
We conclude with a discussion of the Weyl formula.

\subsection{Classical Mechanics}

We study the cardioid billiard whose boundary is defined by
the following mapping of the unit circle in the complex plane
\be \label{cardef}
z(\theta) = e^{i\theta} + {1 \over 2}
e^{i2\theta}, \;\;\; \theta \in [-\pi,\pi),
\ee
and is plotted in Fig.~\ref{card}. The angle $\theta$ is defined such
that it changes discontinuously from $\pi$ to $-\pi$ at the cusp.
If the factor of $1/2$ is replaced by a parameter $b$ this
represents a family of billiards introduced in Ref.~\cite{rob} and 
subsequently studied exhaustively \cite{mark,robber,Hayli,Prosen,Bruus,Li}.
For $b<1/2$, Eq.~(\ref{cardef}) is a conformal mapping but is not strictly 
conformal for $b=1/2$ since the derivative of $z$ with
respect to $\theta$ vanishes at $\theta=\pi$. In practice, this does not matter
and the algorithm introduced in \cite{Prosen} to find the quantum
eigenvalues still applies and was used by us.  For $b>1/2$ the curve
crosses itself near $\theta=\pi$ and the billiard is not well defined. 
We note that the quantum behaviour of the cardioid was recently studied in
Ref.~\cite{backer}. The dynamics in the billiard consists of free
motion within the domain followed by equal angle (specular) reflections
at the boundary.  Trajectories which strike the vertex are not defined
but these are of measure zero. Motion in the cardioid has been proven by 
Markarian to be ergodic \cite{mark}. It is 
similar to the Bunimovich stadium \cite{bun} in that it is defocusing.
Defocusing means that each point on the billiard has positive 
curvature so that parallel rays striking the boundary are initially focused.
However, the billiard geometry is such that the trajectories typically
diverge even more
after the focal point resulting in a net defocusing.  It is this mechanism
which leads to the average divergence of trajectories and to chaos.
In contrast, a dispersing system such as the Sinai billiard \cite{sin} 
has negative curvature so that initially parallel rays
striking the boundary are immediately dispersed.

The curvature of the cardioid is
\be \label{curv}
\kappa(\theta) = {3 \over 4} \sec\left({\theta\over 2}\right),
\ee
which is positive for all $\theta$, as in a circle. For 
$b$ somewhat less than 1/2, the region near $\theta=\pi$ is dispersing rather
than defocusing. Billiards with
mixed focusing properties like this are difficult to analyze mathematically 
\cite{mark,bun2}
and it is for this reason that Markarian's proof works only for the cardioid.
In cartesian coordinates near the cusp the billiard boundary satisfies
the equation 
\be \label{bbound}
y \approx \pm {1\over 2} \left(-(2x+1)\right)^{3/2}.
\ee
Consequently there is a cusp at $x=-1/2$ which locally looks
like a half plane extending to the left. This is an example of
a vertex singularity so that the analysis of the previous section applies.

We found periodic orbits numerically by using the principle of least
action. For an arbitrary periodic orbit the number of intersections with the
boundary was specified and the intersection positions 
were varied until a local minima of the total orbit length
was found. Diffractive orbits were found the same way
but with the constraint that one of the intersections was at the cusp.
Various geometric orbits are shown in 
Fig.~\ref{geoorbit}. The label of each orbit includes the number of 
intersections
and also a letter index to further distinguish them. We describe a better
naming system below. The asterix designates self-dual orbits as defined below.
In Fig.~\ref{diforbit} we show various diffractive orbits. The naming scheme
is similar to before, the number gives the number of bounces --  not counting
the cusp. The primes indicate diffractions, ie.\ the number of encounters
at the cusp. Orbits 2a', 3a'' and *4a'' have arrows
drawn to indicate the scattering directions at the cusp. The last two have
two diffractions and 3a'' is seen to be a composition of 1a' and 2a'

The three orbits *6b, *8b, and *10b are geometric and reflect specularly near 
the cusp, contrary to appearances.
Also, 4a misses the cusp and is geometric in contrast
to 4a' which hits the cusp and is diffractive. 
In Fig.~\ref{pruorbit} we show examples of pruned orbits \cite{prune}
of both the geometric and the diffractive kind. These have an
index p in their label to indicate that they are pruned.
It can be seen that the orbits are related pairwise;
orbit 10p' appears to be composed from 5p and 5a' and similarly for 12p' and
14p'. This is a feature we discuss below.

\subsection{Symbolic Dynamics}

Symbolic dynamics \cite{symbprune} 
is the partitioning and labeling of topologically
distinct regions of phase space. 
Because of the reflection symmetry of the problem, we can discuss the
dynamics either in the full domain or in just half of it. The half domain,
also called the fundamental domain, has dynamics
which are the same as in the
full domain but with a reflection at the symmetry axis.
We will show that the full and fundamental
domains have distinct but closely related symbolic dynamics.
We begin with a discussion of the symbolic dynamics of the geometric orbits.

For an arbitrary, time-reversal invariant system with a reflection symmetry,
all orbits belong to one of five classes.
In principle there can be boundary orbits which 
lie directly on the symmetry axis; in this example there happen to be no
such geometrical orbits.
The other four possibilities are (i) symmetric and self retracing,
(ii)  symmetric but not self-retracing, (iii) self-retracing but not
symmetric, and (iv) neither self-retracing nor symmetric. These occur
with multiplicities 1, 2, 2 and 4 respectively and examples are
*4b, 3a, 6c, and 7b. Although (iv) is the
most generic possibility, we did not find many examples of it among
the shortest orbits. It is typical that the shortest periodic orbits
are special \cite{bent,PCBE} and that 
generic ones begin to appear only for longer lengths. Orbits of classes
(ii)-(iv) behave identically in the half domain as in the full domain.
Orbits of class (i), the so-called self-dual
orbits, must be treated more carefully because in the fundamental
domain they are periodic in $T_\gamma/2$ as well as in $T_\gamma$.

We start by discussing the symbolic dynamics of the full domain.
Recalling that each point on
the boundary of the billiard is labeled by an angle $\theta$ in 
Eq.~(\ref{cardef}), we assign a trajectory the symbol ``$+$'' every time it 
has a reflection
which increases $\theta$ (counterclockwise) and the symbol ``$-$''
every time it has a reflection which decreases $\theta$ (clockwise). 
At the cusp the angle $\theta$ changes discontinuously by $2\pi$, so it
defines what we mean by an increase or decrease of angle.
Since we allow no geometric orbit to hit the cusp, it is not a problem
that the sign of $\theta$ is not defined there.
This is a general property of dynamical systems - the symbolic dynamics is
often conveniently described with reference to a discontinuity 
\cite{symbprune}.

As an example, consider orbit 3a. When going in the counterclockwise sense, it
has the symbol sequence $+-+$. Its time reversed partner, which is distinct in
the full domain, has the sequence $-+-$. 
We are free to start counting symbols anywhere on the orbit so that
any cyclic permutation of a symbol sequence describes the same
orbit and is not distinct. If $W$ is
the symbol sequence of an orbit which is symmetric under reflections 
(such as 3a) then $W=W^*$, where $W^*$ is obtained by reversing the order
of the symbols. For example, $+-+$ reads the same left to right as right to 
left. Self-retracing orbits of length $2n$ (they must be even)
have symbol sequences with the structure $W=A\tilde{A}$ where $A$
is some sequence of length $n$ and $\tilde{A}$ is obtained from $A$ by
reversing the order and every sign. For example, the symbol sequence 
for the self-retracing orbit 6c is $--+-++$ for which $A=--+$ and 
$\tilde{A}=-++$. A 
self-dual orbit is both symmetric and self-retracing and therefore its symbol 
has both properties. An example is *4b whose symbol sequence
is $++--$ (this satisfies the property of being symmetric if 
one makes use of the freedom to cyclically permute the symbols.)

The symbolic dynamics in the fundamental domain are defined
by looking at the symbol sequence of an orbit in the full domain and assigning
a 0 if two adjacent symbols are the same and a 1 if they are not.
For example, orbit 3a, which is labeled $+-+$ in the full 
domain is 011 in the fundamental domain. The time
reversed orbit in the full domain, $-+-$, is also $011$ in the
fundamental domain. This is consistent since in the full domain they are
distinct and should have separate symbols while in the fundamental domain 
they are not distinct and should have the same symbol.

Note that a self-dual orbit in the full domain has a symbol sequence in the
fundamental domain which repeats itself, for example *4b has the sequence
$0101=(01)^2$. Therefore a self-dual orbit, when mapped onto
the fundamental domain, is
the double repeat of a shorter orbit. Every odd multiple of this
shorter orbit will be present in the fundamental domain but not in the full
domain and this is apparent in the symbolic sequences. Any sequence in which 1 
appears an odd number
of times can not be periodic in the full domain and so corresponds to 
a self dual orbit. Any
orbit which is symmetric in the full domain is self-retracing in the half
domain.

We show in Table~I, the symbols of all orbits up to length 4 as 
measured in the fundamental domain. There is no fundamental
orbit 0, this is reminiscent of the co-linear helium problem \cite{tanner}, 
among others. The pruned family 5p,6p,7p... means that there are no
orbits with the symbol sequence $0^n11$ with $n>2$. On the other hand, there is
an accumulation of whispering-gallery-like orbits labeled 
5a,6a,7a... whose symbol sequences have the form $0^n101$ and whose
lengths accumulate to $L=12$ as $n\rightarrow\infty$.
There are also orbits of the form $01^n$; for $n$ even they are
3a,5b,7c... while for $n$ odd they are self-dual and
are 4b,8b,... We believe that both series exist for any $n$.

The symbolic dynamics of diffractive orbits 
is clearest if one think of the cusp as not being part of the boundary but
rather being a means of getting from one point on the boundary to
another. We introduce a symbol $d$
which represents a path between two points on the boundary which goes via
the cusp and keep $+$ and $-$ as defined above.  Therefore, orbit 2a'
has two boundary intersections and has a symbol sequence
$d+$ or $d-$ depending on the sense of the rotation.
Symmetric diffractive orbits labeled $dW$ satisfy $W=W^*$ while self
retracing orbits have the structure $W=A\tilde{A}$, as for geometric orbits.  

The rule for the symbolic dynamics in the fundamental domain is again found 
by looking at the word in the full domain. A symbol $d$ maps to a $d$, while
a $+$ or $-$ maps to a 0 or 1 depending on whether the next non-$d$ symbol
is the same or different.  For example, in the half domain, 2a' has the 
symbol $d0$. In the fundamental domain the two rotation senses of 2a' are not
distinct and it is consistent that there is only one symbol sequence.
The geometric orbits discussed above can be subsumed into
a larger ternary alphabet in which they are the subclass with no $d$ in their
symbol sequence. 

As before, any symbol sequence with an odd number of 1's must correspond to
half of a self-dual orbit. However, self-dual orbits with one diffraction
have the special property of being
geometrically identical to a non-self-dual diffractive orbit 
and we refer to the pair as complements. An example of this is the self-dual
orbit *4a'' which is a perfect overlap of
orbit 2a', the only difference being that the first backscatters at the cusp 
while the second forward scatters.
Consequently, *4a'' has twice the length of 2a' and suffers two diffractions 
rather than one. This is general, the self-dual orbit
always leaves the cusp at an angle which is the negative of its complement
and then follows a trajectory which is simply the
reflection of its complement.
If an orbit in the full domain has the symbol sequence $dW$ then its self
dual complement has the symbol sequence $dWdW'$ where $W'$ is
defined such that $WW'$ satisfies the self-dual property. For example,
*4a'' has the symbol
sequence $d+d-$ and $+-$ is clearly self-dual. In the
fundamental domain the symbol sequence of a self-dual 
orbit is found from its complement by switching the symbol immediately
before the $d$. For example, the symbol sequence of 3b' is $d11$ while that
of its self dual complement (not shown) is $d10d10$ 
(recall that due to the cyclic symmetry,
the last character in these sequences is ``before'' the $d$).
In Table~II we show the labels of all singly diffractive orbits up to 
length 4. Note that the symbols $d101$ and $d011$ are simply time reversed
copies of each other and contribute equally to the trace.


To find the symbol sequence of any multiply diffractive orbit, we use
the same rule. For example, the doubly diffractive orbit 3a'' starts at the 
cusp, travels along 1a' back to the cusp,
diffracts onto the 2a' orbit, travels along 2a' back
to the cusp and finally diffracts 
onto the 1a' orbit in the original direction. Its symbol sequence is
$d0d$. 
Because there is only one vertex, the only multiply diffractive orbits are
compositions of singly diffractive ones. 
More possibilities would exist if there were more than one vertex.

If all possible symbol sequences were realised as
orbits, the number of singly diffractive orbits in the fundamental
domain up to length $n$ would grow as $2^n$. 
For geometric orbits with a complete binary alphabet, the number grows
as $2^n/n$.  The factor of $n$ in the denominator is because cyclically 
permuted symbols correspond to the same orbit and should be counted only once.
It appears that there are more singly diffractive orbits of long length and
it is not clear if they ultimately dominate the spectrum. A related issue 
is whether the pruning of geometric and
diffractive orbits is such that the exponential proliferation of the two 
classes of orbits is given by the same exponent.
This is quite likely since very long orbits cover the phase space uniformly
and are therefore susceptible to the same pruning mechanisms as shown
in Fig.~\ref{pruorbit}. These questions will be studied in 
greater detail in a later publication \cite{hbnw} and
here we have just a brief discussion.

The pruning
of geometric and diffractive orbits appear to be strongly correlated. For
example, the pruned diffractive orbit 10p' looks as if it is composed
of the pruned orbit 5p and the diffractive orbit 5a'. 
This is confirmed by looking at the symbol sequences of these three orbits. 
Orbit 10p' has the word $d++++-++++$ which is equal to the composition of
those for 5a' and 5p which are $d++++$ and $-++++$, respectively. 
A similar result holds for orbits 12p' and 14p'. Usually, the existence
of pruning implies problems in the cycle expansion of the zeta function
due to the breakdown of shadowing \cite{symbprune} 
so the fact that orbits and 
their shadows have disappeared together might prove very useful. An example
of non pruned shadowing are 
$5b\Leftrightarrow 2a+3a$ (since $+-++-\; = \; +-||++-$) and
$4c'\Leftrightarrow 2a'+2a$ (since $d+-+ \; = \; d+||-+$).

The symbolic dynamics was useful to us in guessing the topology
of a few missing orbits.  However, we did not make extensive use of it as we 
were
content to know the shortest orbits and these can be found by trial and error.
It is useful, however, in the following discussion of Maslov indices and 
symmetry reduction.
In addition, a project which required knowing many orbits, such as attempting
to find semiclassical approximations to many quantum eigenvalues, would need to
make extensive use of the symbolic dynamics.

\subsection{Maslov Indices}

In this subsection we discuss the Maslov indices which appear in the trace 
formulas (\ref{tf_geom}) and (\ref{tr1}), beginning with
the geometric indices. Starting at an arbitrary point $x$ on the
orbit, the index $\sigma_\gamma$ 
equals the number of caustics $\mu$ plus an index $\nu$ which arises 
on doing the stationary phase integral in the determination of the trace.
Although $\mu$ and $\nu$ depend separately on the point $x$ along the orbit,
their sum does not. In fact, $\sigma_\gamma$ is a canonical invariant
\cite{stevie}  which equals twice the number of times that the stable and 
unstable manifolds
wind in completing one circuit of the periodic orbit. It follows that we can
use any point $x$ on the orbit to calculate $\sigma_\gamma$.

We numerically propagated the $2\times 2$ monodromy matrix $M$ for each
geometric
orbit and counted the number of caustics $\mu$ by the number of times that
one of the off diagonal elements $M_{12}$ changed sign. At the end, we also
found
the value of $(\mbox{Tr}M-2)/M_{12}$; if it is positive $\nu=0$ and
if it is negative $\nu=1$. Doing so, we found the following simple topological
rule; $\sigma_\gamma$
equals the number of reflections or, equivalently, the length of
the symbol sequence of that orbit.
This rule probably arises from the fact that the cardioid
is a purely defocusing billiard so there is, on average, one focus per
reflection.  In the Bunimovich stadium, one finds a similar rule that 
$\sigma_\gamma$ is incremented by one for each reflection off the defocusing
end caps \cite{sccpri}. It is common that such a simple
rule exists and it is usually related in some simple way to the symbolic
dynamics.

For the diffractive orbits $\sigma_\gamma$ is the number of
caustics between successive diffractions.
We found that $\sigma_\gamma$ always equals the number of 
geometric reflections and hence the length of the symbol sequence,
as before. This appealing result implies a unity between the two classes
of orbits.

Every geometric reflection also induces a sign change due to Dirichlet boundary
conditions. We account for this by incrementing $\sigma_\gamma$ by 2 at
every reflection, so that in total $\sigma_\gamma=3m_\gamma$ where 
$m_\gamma$ is the number of reflections experienced by orbit $\gamma$.
For Neumann boundary conditions this is not necessary and 
$\sigma_\gamma=m_\gamma$. Consistent results, specialised to the half
cardioid, were obtained in Ref.~\cite{backer}.

\subsection{Symmetry Decomposition}

The billiard has a reflection symmetry $C_2$ and consequently all
quantum states can be classified as even or odd. The trace decomposes
as 
\be \label{trdec}
g(E) = g_+(E) + g_-(E).
\ee
We can separately find $g_\pm (E)$ by studying the dynamics in the 
fundamental domain and from them the densities $\rho_{\pm}(E)$. The
behavior of the geometric trace formula (\ref{tf_geom}) under symmetry 
decomposition is a well studied problem \cite{bent,PCBE,robbins,creagh}
and here we review the results which are relevant to us.
Thereafter, we discuss the decomposition of the diffractive trace
formula, which is slightly different.

The non-self-dual orbits have multiplicities of either 2 or 4 in the full
domain and their amplitudes are divided equally between
the two parity classes. Self-dual orbits require more care. Half of such an
orbit, being periodic in the fundamental domain, contributes to the the
separate traces as follows. Its period,
stability and Maslov index are all half of the full orbit and its stability
is the square root. In addition the $\mp$ factor in the denominator of 
Eq.~(\ref{tf_geom}) is replaced by $\pm$. Finally, there is a group
theoretical 
weight of $\pm$ corresponding to the even/odd parity. This last factor ensures
that the contribution of this half orbit identically cancels when we evaluate
the sum
$g(E) = g_+(E) + g_-(E)$. This is consistent since the half orbit is not a 
periodic orbit of the full domain and should not affect the total density
of states. The double repeat of a half orbit is the full orbit and its 
amplitude is divided equally between the two parities.

We next discuss what happens to the diffractive orbits in the presence of the
symmetry. There is a diffractive boundary orbit; it contributes only to the
even spectrum \cite{ndw1,ndw2,bent}, unlike a geometrical boundary orbit
which contributes to both. The distinction can
be traced to the fact that the diffractive
Green function (\ref{g_diff}) is multiplicative in the direct Green 
functions.
For the other diffractive orbits, we recall the previous discussion
that each of them has a self dual complement of twice the length.
The only difference between
the orbits in the fundamental domain
is that they have different diffraction constants. If one has the 
diffraction constant $d(\theta,\theta')$ the other has the
diffraction constant $d(-\theta,\theta')$. Since they are in all other 
respects identical, we can include both of them by defining
separate diffraction constants for the even and odd parities, namely
\be \label{dpar}
d_\pm(\theta,\theta') = d(\theta,\theta') \pm d(-\theta,\theta').
\ee
The different sign for the two cases is the same group theoretical weight
mentioned for self-dual geometric orbits.  

As mentioned above, for the half plane and for Dirichlet boundary conditions,
$d(-\theta,\theta')=d(\theta,\theta')$ so we have
\be \label{odir}
d_+(\theta,\theta') = 2d(\theta,\theta') \;\;\;\;\;\;\; 
d_-(\theta,\theta') = 0.
\ee
This implies that the odd spectrum is completely
insensitive to the existence of diffraction. For wedges which are not 
half-planes, Eq.~(\ref{odir}) is not true but it will still be true that
Eq.~(\ref{dpar}) will cause the two parities to be affected differently. If
we study the billiard with Neumann boundary conditions we reach the opposite
conclusion. First, the boundary orbit has zero amplitude since 
Eq.~(\ref{dspec}) implies that $d(0,0)=0$ in
that case. Also, $d(-\theta,\theta') = -d(\theta,\theta')$ for Neumann
boundary conditions so 
that
\be \label{edir}
d_+(\theta,\theta') = 0 \;\;\;\;\;\;\; 
d_-(\theta,\theta') = 2d(\theta,\theta').
\ee
Therefore, the odd spectrum would have diffractive peaks and the even spectrum
would not.

\subsection{Weyl Formula}

The Schr\"{o}dinger equation for a billiard reduces to the Helmholtz equation
\be \label{helm}
\left(\nabla^2 + k^2\right)\psi(r) = 0
\ee
with Dirichlet, Neumann or mixed boundary conditions on the boundary of the
domain. Finding the eigenvalues $k_n^2$ leads to the density 
of states $\rho(k) = \sum_n\delta(k-k_n)$. One commonly decomposes this into
a smooth part and a fluctuating part
\be \label{decomp}
\rho(k) = \bar{\rho}(k) + \rho_{\mbox{fl}} (k).
\ee
These terms have distinct classical interpretations. The first term,
commonly called the Weyl term, is related to the 
geometry of phase space, such as the area, perimeter, curvature and other
properties of the
billiard boundary. The second term is given by the dynamics as encoded in
the trace formulae (\ref{tf_geom}) and (\ref{tr1}). Actually, each
term of Eq.~(\ref{decomp}) is an asymptotic expansion in powers of $1/k$
\cite{berhow}. To date, the first 16 terms of the expansion of $\bar{\rho}(k)$
have been calculated but here we content ourselves with the first three.
Additionally, the first corrections to $\rho_{\mbox{fl}}(k)$ have also
been determined \cite{gasal,gabor} but we do not consider them here.

Instead of $\bar{\rho}(k)$, one often refers to the spectral staircase function
$\bar{N}(k)$ of which $\bar{\rho}(k)$ is the derivative.  Its expansion 
for Dirichlet boundary conditions is (see for example Ref. \cite{berhow})
\be \label{strfn}
\bar{N}(k) \approx {A \over 4\pi}k^2 - {L \over 4\pi}k + C - \cdots
\ee
$A$ is the area of the billiard, $L$ is the length of the perimeter,
and $C$ is related to 
the curvature and to corners by
\be \label{ccurv}
C = {1\over 24\pi}\sum_i {\pi^2-\theta_i^2 \over \theta_i}
+ {1 \over 12\pi} \int\kappa(s)ds.
\ee
The sum is over angles in the billiard boundary; we have one
angle of $2\pi$. The integral gives the total curvature over the boundary
of the billiard.  Note that Eq.~(\ref{strfn}) also applies to billiards
with Neumann boundary conditions if we multiply every other term by $-1$.
For the full spectrum, we find
$A=3\pi/2$, $L=8$ and $C=3/16$.  We can also use Eq.~(\ref{strfn}) for
the odd spectrum by taking the billiard domain to be the fundamental 
domain.
We then have $A=3\pi/4$, $L=6$ and $C=3/16$.
The difference between the total spectrum and the odd spectrum must correspond
to the even spectrum, so for it we have
$A=3\pi/4$, $L=2$ and $C=0$. The symmetry decomposition of the Weyl
formula for billiards was discussed in much greater generality in 
Ref.\cite{pavloff}. These results for the cardioid billiard were also
obtained in Ref.~\cite{backer}.

Comparing the exact staircase functions with the approximation 
(\ref{strfn}) is a useful check that there are no missing levels. 
In addition, the point where there the curves start to deviate is a useful
criterion for establishing when the numerical eigenvalues are no longer
reliable. The Weyl formula is also needed to compare the Fourier transforms of
the data and the trace formulae in Section~IV and also to renormalise the 
spectrum for the statistical analysis of Section~V.

\section{Numerical Results}

In this section we present comparison between the exact spectra and the results
of periodic orbit theory.  We first do this by directly comparing the
Fourier transforms of the exact spectra and the trace formulae in the 
reciprocal space of orbit lengths $L$. Overall, there is good agreement
and we successfully reproduce geometric, diffractive and doubly diffractive
peaks as well as the interferences among them. However,
there is a region of $L$ which is not well reproduced for reasons we explain.
We also find that for other
regions of $L$, the exact diffractive peaks have magnitudes larger by a few
percent than what we expect. We explain this by
recalling that the cusp is not a perfect vertex.  Finally, we
use the sum over periodic orbits to find the first few quantum eigenvalues and
find the odd result that including diffractive orbits seems to shift them
only slightly.

\subsection{Fourier Transforms}

Using the algorithm of Ref.~\cite{Prosen} we truncate the Hilbert
space to contain only the lowest 6600 states and calculate the lowest
1000 eigenstates of each parity with an accuracy better than 0.001
times the average level spacing. It is these 2000 states we use in the
analysis. For a precise comparison, it is best to work with 
$\rho_{\mbox{fl}}(k)$ which is obtained from subtracting the Weyl term 
$\bar{N}(k)$ of Eq.~(\ref{strfn}) from
the exact density of states.  We obtain its Fourier transform as
\be \label{fft}
F(L) = \int_{-\infty}^{\infty}dk \: w(k)[\rho(k)-\bar{\rho}(k)]e^{ikL}
\ee
where $w(k)$ is a window function. We chose to use the 3-term Blackman-Harris
window \cite{harris}
which gives a good compromise between narrowness of peaks and smallness
of side lobes. This is defined as
\be \label{window}
w(k) = \sum_{j=0}^2a_j\cos\left(2\pi j{k-k_0 \over k_1-k_0}\right).
\ee
where $(a_0,a_1,a_2)=(0.42323,-0.49755,0.07922)$. This function goes
smoothly to zero at $k_0$ and $k_1$ which we choose as the
of the first and last eigenvalues in our spectrum.

We apply the same Fourier transform to the trace formulae Eqs.~(\ref{finally})
and (\ref{rhog}) to obtain the semiclassical approximation 
$F_{\mbox{sc}}(L)$. We included all relevant period orbits with $L < 11$;
the most was for the even spectrum which had 38 orbits, including halves of
self-dual orbits and multiple repeats of 2a. The results up to 
$L=10.7$ are shown in Fig.~\ref{ft1} for the even, odd and combined spectra.
Other than the 
region around $L=7.5$ which we discuss later, the agreement is very good. This
was also observed in Ref.~\cite{ps} but here we are also verifying the
geometrical factors $F$ in Eq.~(\ref{finally}). Note
that there are no diffractive peaks in the odd spectrum as we argued above.
The geometric peaks in the even and odd spectra near $L=2.5$,
$L=4.7$, $\cdots$ are halves of 
self-dual orbits and are absent in the full spectrum.

We stress that the relative heights of the
diffractive peaks are artifacts of the range of the quantum spectrum we 
choose to consider and should not be used to estimate the relative weight of
these orbits in determining quantum eigenvalues. The reason is that they are
suppressed as $1/\sqrt{k}$ and contribute more and more weakly to the energetic
states. However, their effect in the ground state region can be quite large --
a possibility we explore later.

We now turn our attention to the large discrepancy near $L=7.5$. There are two
distinct reasons why the trace formulas fail there. The first is that orbits 
4a, 4a' and *10b are close in configuration space and interfere
differently than we have assumed up to now. The second reason is
that the cusp is only approximately a half-plane vertex.
To understand the first point, we appeal to the calculation which gives the
diffraction constant. Sommerfeld \cite{somm}  
showed that an incoming plane wave is broken up by a half plane vertex into 
three components; a plane wave which continues in the original direction,
another plane wave coming from reflection off one face, and a third component 
which he identified as the diffracted field.
The diffracted field is asymptotically an outgoing circular 
wave of the form $f(\theta)\exp(ikr)/\sqrt{kr}$ and the diffraction constant 
$d$ is proportional to $f(\theta)$.  However, in the directions close to 
the two out going plane waves, the diffracted wave takes longer and longer to 
obtain its
asymptotic form and exactly in the plane wave directions it never does.
It is this attempt to connect to the incorrect asymptotic form in these two
directions which leads to the divergence of Eq.~(\ref{dcons}) when
$\theta\pm\theta'=\pi$. Ideally, one would like to have a uniform approximation
to cover all ranges of $\theta$ and $\theta'$. In the case discussed here, we
will find that the approximation improves as $k$ increases.

The second reason for the failure of the trace formula near $L=7.5$ is
that the billiard domain departs relatively rapidly from its local
half plane geometry at the cusp so that the diffraction constants derived
assuming a half plane may not be appropriate. In fact, this approximation is
fine for orbits which come in from the right, such as 1a', since they are never
close to the faces of the cusp. However, orbits entering the cusp from the
left, such as 5a', are very sensitive to this approximation. 
The curvature of
the boundary means that such orbits are not as far from the boundary as 
is assumed in the calculation of the diffraction constant so that the 
Dirichlet boundary conditions cause more suppression than is accounted for.

To understand this
another way, note that for any finite wave number $k$ there is not 
infinite spatial resolution so the cusp appears as a finite 
angled wedge. However, a finite angled wedge has different diffraction
constants. Therefore, it is reasonable to suppose that at small wavenumbers the
lack of resolution, inherent in the finiteness of $k$, manifests itself as
inaccuracies of peak amplitudes in the Fourier spectra.
We explored this effect for orbit 1a' by looking at two
windows of $k$. For this study, it is best to use the combined spectrum 
because then there is no interference from the self-dual orbits 2b and 4b.
We are extremely sensitive to errors because we are interested
in the difference between two small peaks. Accordingly, we use just the
lowest 285 states which are extremely accurate 
(approximately 140 of each parity) which we divided into
two windows of approximately equal extent in $k$ (2.01-26.32 and 13.09-39.39
respectively with means $\langle k\rangle = 14.17$ and 26.24.) 
The comparison between the numerics and the trace formula
\ref{finally} is shown in Fig.~\ref{2a}. In both cases, we find that the exact
spectrum has a peak which is slightly larger than predicted but that the
discrepancy is larger by about 50\% for the first window.

This strengthens our argument that the large discrepancies around $L=7.5$ in
Fig.~\ref{ft1} arise because the orbits responsible for those peaks are  
sensitive to the fact that the cusp is not a true half plane. We argue that
because they approach the cusp from the left they are far more sensitive
to this than orbits approaching from the right. Furthermore, such orbits 
are probably sensitive to the entire curved geometry in the 
neighborhood of the cusp.

The fact that there is structure in the odd spectrum with lengths the same
as for diffractive orbits (particularly visible near $L\approx 7.5$, an effect
also observed in Ref.~\cite{backer}) is not
unexpected. In one of their seminal papers, Balian and Bloch \cite{balblok}
argued that for a billiard domain, with a discontinuity in the n'th
derivative of the boundary, there exists contributions to the density
of states of order $\hbar^{n/2}$. For sharp corners, we have $n=1$ and this
conforms to the previous discussion. For discontinuous changes in the curvature
we have $n=2$. The second case applies
here for the odd spectrum since the half cardioid has continuous slope
but its curvature changes discontinuously from zero to infinity at the
vertex. Normally, this structure is too small to be visible but we suppose
that the geometrical considerations which amplify the diffractive
peaks in the even spectrum have a similar effect on the higher order peaks
in the odd spectrum.

Another possibility for contributions due to a change in curvature is 
discussed in Refs.~\cite{ssclgr,alonso} in the context of the Bunimovich
stadium \cite{bun}. It was found
that although one can unambiguously continue a classical trajectory
which encounters the discontinuity, the stationary phase
integral to evaluate its contribution to $\rho(E)$ requires more care. This
is because the character of the motion is different on each side of the orbit.
This applies to orbits which reflect specularly at the vertex
and so are geometrical - not diffractive.

Finally, we discuss the small peak near $L=9.8$ corresponding to orbit 3a''. 
We use Eq.~(\ref{finally}) with the appropriate diffraction constants. 
To eliminate interference from the nearby self-dual peak 8c we 
analyze the combined spectrum.
Because this is such a small peak, it is particularly sensitive
to small backgrounds.  One source of background is slight errors
in the determination of the exact quantum energy levels; such errors manifest
themselves as weak oscillations in the Fourier transforms. 
For this reason, we used only the lowest 1000 levels. In
Fig.~\ref{2a3b}, we show  
both the absolute value of the Fourier transform and its real part. 
The discrepancy, which grows towards the right of the figure, comes from
a slight error in the peak amplitude of 4b'.

\subsection{Recovering Quantum Eigenvalues}

The trace formulae Eqs.~(\ref{finally}) and (\ref{rhog}) can be used
to find quantum eigenvalues. Here we do this 
with and without the diffractive orbits to see what effect they have on the
determination of the eigenvalues. We simply summed over the orbits used to
obtain Fig.~\ref{ft1}. Because of the problems with the diffractive orbits
5a', 6a' ... we ignored that series of orbits. Additionally, we made an
approximate fix for small $k$ of the orbits 4a, 4a' and *10b  by ignoring
4a' and *10b and halving the amplitude of 4a. This simulates the effect of
the cusp which has the approximate effect of halving the domain of the
stationary phase transverse integral used to derive the geometrical
trace formula. The result for the
even spectrum is shown in Fig.~\ref{gf1} with and without diffractive orbits. 
The peaks are identified as corresponding to quantum eigenvalues.
We observe that including diffraction helps to resolve the peaks but barely
changes their positions. The positions are compared in Table~\ref{eigdat}.
The discrepancy is typically .06 compared to a spacing of about 0.80.
Including the diffractive orbits changes the peak positions only by about .02
and not necessarily by the correct sign.

This is a strange result because the amplitude of the geometric and diffractive
contributions to the density of states are comparable, as shown in the right
half of Fig.~\ref{gf1}. We saw in the previous section that the eigenvalues 
do contain information about
diffraction since there were diffractive peaks in the Fourier transform so it
must be true that the diffractive orbits have some
effect on the eigenvalues. To unravel this paradox it is probably best to
calculate with zeta functions rather than traces, which we will do
in a later publication \cite{hbnw}.
For now we note that the wedge billiard was also successfully quantised using
only geometric orbits although it too has a diffractive vertex \cite{wdglit}.

For completeness, we show in Fig.~\ref{gf2} the results for the odd spectrum
and the results are also enumerated in Table~\ref{eigdat}.  Now only
geometric orbits contribute. The agreement is somewhat better; with the 
exception
of the third state, which is not well resolved in any case, 
the differences between the
peak positions and the quantum eigenvalues are typically about .04. The
only qualitative difference between the two parities is the 
diffraction so the difference in accuracy is presumably diffraction related.

\section{Spectral statistics}

It is by now well known, if not well understood, that the statistics of 
eigenvalues of chaotic systems \cite{boh} follow closely the predictions of 
random matrix theory (RMT) (for a review see Ref.~\cite{mehta}). This has been 
confirmed in many examples including the conformal mapping of the circle 
\cite{Prosen} of which the cardioid is a limiting case.
Nevertheless, we repeat this here to see if the diffraction has any effect
on the statistics. The fact that the diffraction is almost completely limited
to the even spectrum means that we can study its relative effect by comparing
the results for the even and odd spectra. A similar study was already
reported in Ref.~\cite{backer} for the half cardioid but without an
explicit comparison of the odd and even spectra to explore possible 
diffraction effects.

We first display the spacing distributions in Fig.~\ref{PofS}. This is
the distribution 
of the spacings $s$ between adjacent energy levels measured in units of the
local mean level density as found from the Weyl formula Eq.~(\ref{strfn}). 
The Gaussian Orthogonal Ensemble (GOE) is a random matrix ensemble which
predicts a spacing distribution very close to
\be \label{psgoe}
P(s) = {\pi \over 2}s\exp(-\pi s^2/4).
\ee
Both spectra are
observed to be consistent with that limit and with each other so diffraction
appears to have no significant effect on the spacing distribution.

The spacing distribution is a measure of short range correlations. A statistic
commonly used to probe the long range correlations is the spectral rigidity
$\Delta_3(l)$. Here $l$ refers to distances in the spectrum measured 
in units of the mean level
spacing. (We use a small letter $l$ to avoid confusion
with the periodic orbit lengths which we referred to with a large $L$.) 
This statistic measures the average $\chi^2$ deviation of
the staircase function from a local straight line fit over a window of
length $l$; the average being taken as the window is moved through
the spectrum. The GOE formula is given as a complicated integral 
representation but for large $l$ is nearly $(\log l)/\pi^2$.
Chaotic systems follow the GOE result \cite{berry}
but at some point begin to deviate
from it and finally saturate for arguments larger than about
$l_{\mbox{max}}=2\pi\langle\rho(k)\rangle/L_{\mbox{min}}$ 
where $\langle\rho(k)\rangle$ is the average 
level density and $L_{\mbox{min}}$ is the length of the 
shortest periodic orbit. The shortest periodic orbit in the cardioid
is 2a which has length 2.60 in the fundamental domain. The average
density of states in the range considered
is 20.0 so that $l_{\mbox{max}}=48.3$.

In Fig.~\ref{Delta3}, we show the $\Delta_3(l)$ results for the even
and odd spectra.  
We also show the saturation value of 0.265 
which we found numerically. As can be seen,
$l_{\mbox{max}}$ is a good approximation to where the saturation begins.
The results are significantly different from the GOE prediction for values of
$l$ larger than about 7. In comparing the two data sets, we observe that
the odd spectrum has values which are consistently larger than the even
spectrum for $0<l<17$. The difference is typically about 0.01.
To determine if this is significant we need to compare the difference
to the typical variance in $\Delta_3(l)$. If we assume that each range of
length $l$ used in determine $\Delta_3(l)$ is statistically independent, we
find that the typical variance is 0.006. This assumption is problematic since
we know that there are strong correlations in the spectra. If instead we
determine the variance by finding the effect of removing selected levels, then
the typical variance is 0.02 \cite{lindemann}. 
In either case, we conclude that if there are 
significant deviations between the two spectra, we are not sufficiently 
sensitive to resolve them with confidence. This conclusion is in agreement 
with the results of the wedge billiard \cite{wdglit} and of pseudo-integrable
billiards \cite{debu}. We can increase the statistical
significance of this statement by finding more eigenvalues.
It would also be interesting to study this question analytically by
extending the Berry's original semiclassical calculation
of spectral rigidity \cite{berry} to include the effect of diffractive orbits.

\section{Conclusion}

In this paper, we have derived a trace formula for periodic orbits
diffracted by vertices. The presence of the
diffractive orbits causes additional structure in the quantum spectra
of Hamiltonian systems. This structure is suppressed relative to the
contributions from geometric orbits. The diffractive trace formula
has a very similar structure to the
trace formula for geometric orbits and from it we can find a zeta
function in close analogy to the zeta function for geometric orbits.
An important difference in the structure of these functions is that
the diffractive zeta function involves only one product. Multiplying
these zeta functions gives the total zeta function which will probably
provide the cleanest method of finding the semiclassical eigenvalues.

We specialised the discussion to the example of the cardioid billiard
possessing a cusp at the boundary which is locally a half-plane vertex. 
There is overall good agreement between the Fourier transform of the
exact spectrum and that of the trace formulae and we successfully
reproduced geometric, diffractive and doubly diffractive peaks as well
as the interferences among them. However, there is a
region of $L$ which is not well reproduced.  Two reasons for this disagreement
are that for certain choices of angles the 
diffraction picture breaks down and because
the cusp is not a perfect half-plane vertex.

There exists a symbolic dynamics which includes the
periodic diffractive orbits in a natural way by
inclusion of one more symbol in the alphabet. This leads to a simple result
when discussing the symmetry reduction. For every non self-dual diffractive
orbit, there exists a complementary self-dual one. These interfere so that
the diffraction affects only the even spectrum and leaves the odd spectrum
alone. We used the sum over periodic orbits to find the first
few quantum eigenvalues with an accuracy of a few percent. We found
the puzzling result that including diffractive orbits seems to have very 
little effect. This issue will be addressed in a later publication.


In the last section we studied the level statistics of the even and odd
spectra separately.
Comparing the results from the two spectra is a probe of the effect of the
diffractive orbits. This is because the odd spectrum has essentially no 
contribution from diffraction. We found no significant differences in either
the spacing distributions or $\Delta_3(l)$ although we can not rule out the
possibility that such differences might become apparent if we included 
more states.

\section{Acknowledgements} 
We gratefully acknowledge illuminating discussions with Debrabata Biswas,
Stephen Creagh, Predrag Cvitanovi\'{c} and Gregor Tanner.
H.B.\ and N.D.W.\ were supported by the European
Commission under grant nos.\ ERBCHBGCT 930511 and 
ERBCHBGCT 930407 respectively.

Note added in proof: After submission of this paper, a preprint appeared
\cite{backer2} in which the symbolic dynamics of the cardioid are extensively
explored. The results of that paper are consistent
with those discussed here.

\begin{table} 
\begin{tabular}{crr}
orbit & full domain & fundamental domain   \\
\hline
2a & $+-$        & $1$                     \\
4b & $+-+-$      & $0101      = (01)^2$    \\
3a & $+-+$       & $011$                   \\
6b & $+++---$    & $001001    = (001)^2$   \\
4a & $+++-$      & $0011$                  \\
8b & $++++----$  & $00010001  = (0001)^2$  \\
8c & $+--+-++-$  & $10111011  = (1011)^2$  
\end{tabular}
\caption{Some geometric orbits and their symbols.}
\label{geomsym}
\end{table}

\begin{table} 
\begin{tabular}{crr}
orbit & full domain  & fundamental domain\\
\hline
1a'  &        $d$ &                   $d$ \\
2a'  &       $d+$ &                  $d0$ \\
4a'' &     $d+d-$ &       $d1d1 = (d1)^2$ \\
3a'  &      $d++$ &                 $d00$ \\
3b'  &      $d-+$ &                 $d11$ \\
4a'  &     $d+++$ &                $d000$ \\
4c'  &     $d+-+$ &                $d110$ \\
4b'  & $d+--$     &                $d101$ \\
\end{tabular}
\caption{Some diffractive orbits and their symbols.}
\label{diffsym}
\end{table}

\begin{table}
\begin{tabular}{cccc}
State & Exact & Geometric & Geometric and \\
number & eigenstate & orbits & diffractive orbits \\
\hline
1 & 2.010 & 1.956 & 1.943 \\
2 & 3.331 & 3.434 & 3.406 \\
3 & 4.169 & 4.053 & 4.070 \\
4 & 4.686 & 4.629 & 4.618 \\
5 & 5.292 & 5.238 & 5.248 \\
\hline
1 & 3.020 & 2.979 & ---\\
2 & 4.160 & 4.134 & ---\\
3 & 5.162 & 4.999 & ---\\
4 & 5.558 & 5.356 & ---\\
5 & 6.174 & 6.133 & --- 
\end{tabular}
\caption{Top: even states calculated with and without diffractive orbits.
Bottom: odd states.}
\label{eigdat}
\end{table}

\begin{figure}
\caption{(a) A path connecting the point $x'$ to the point $x$ via the
vertex $\xi$. The angles $\alpha$, $\theta'$, and $\theta$ appear in
Eq.~(\protect\ref{dcons}) defining the diffraction constant.
(b) A schematic diagram of a periodic diffractive orbit with its local 
coordinates.}
\label{wedge}
\end{figure}

\vspace*{-1mm}

\begin{figure} 
\caption{The cardioid billiard. A generic point $z(\theta)$ at the perimeter is
shown as well as the two special points $\theta = 0$ and $\theta =\pi$ 
(the cusp). We restrict $\theta$ to the interval $[-\pi,\pi)$. The
dashed line indicates the symmetry axis.} 
\label{card}
\end {figure}

\vspace*{-1mm}

\begin{figure} 
\caption{Various geometric orbits of the cardioid billiard labeled with their
names and their lengths.}
\label{geoorbit}
\end {figure}

\vspace*{-1mm}

\begin{figure} 
\caption{Various diffractive orbits of the cardioid billiard labeled with their
names and their lengths. We indicate the incoming and outgoing directions
where it is ambiguous.}
\label{diforbit}
\end {figure}

\vspace*{-1mm}

\begin{figure} 
\caption{A family of related pruned geometric orbits and pruned
diffractive orbits labeled with their names and their lengths.}
\label{pruorbit}
\end{figure}

\vspace*{-1mm}

\begin{figure} 
\caption{The solid curves are the Fourier transforms of the exact spectra
and the dashed curves are the approximations from the trace formulas 
Eqs.~(\protect\ref{finally}) and (\protect\ref{rhog}).}
\label{ft1}
\end{figure}

\vspace*{-1mm}

\begin{figure}
\caption{Comparison between the exact result (solid curve) and the diffractive
trace formula (dashed curve) for orbit 1a'. The top box is the window
$2.01\leq k\leq 26.32$ and the bottom box is the window
$13.09\leq k\leq 39.39$.}
\label{2a}
\end{figure}

\vspace*{-1mm}

\begin{figure} 
\caption{As in Fig.~\protect\ref{ft1} but for a range corresponding to
the double diffractive orbit 3a''.}
\label{2a3b}
\end{figure}

\vspace*{-1mm}

\begin{figure} 
\caption{Left: The solid curve is the result of using both 
geometric and diffractive orbits in the trace formula for the even spectrum
while the dashed curve
is from using just the geometric orbits.  The arrows denote the exact
positions of the even states.
Right: A comparison of the magnitudes of the geometric (top) and diffractive 
(bottom) contributions to the density of states.}
\label{gf1} 
\end{figure}

\vspace*{-1mm}

\begin{figure} 
\caption{The odd density of states as constructed from just the geometric
orbits and the arrows denote the exact positions of the odd states.}
\label{gf2}
\end{figure}

\vspace*{-1mm}

\begin{figure} 
\caption{Histograms showing the spacing distributions of the even and
odd spectra. The solid curve is the GOE result.}
\label{PofS}
\end{figure}

\vspace*{-1mm}

\begin{figure} 
\caption{The $\Delta_3(l)$ functions for the even ($\circ$) and the
odd ({\protect\small $\times$}) spectra. The solid curve is the GOE
result and the dashed line denotes the saturation value.} 
\label{Delta3}
\end{figure}

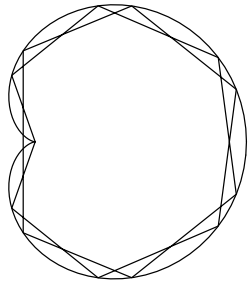
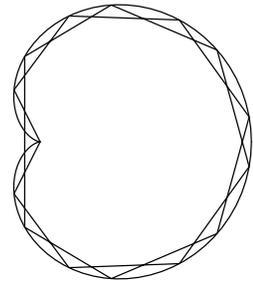
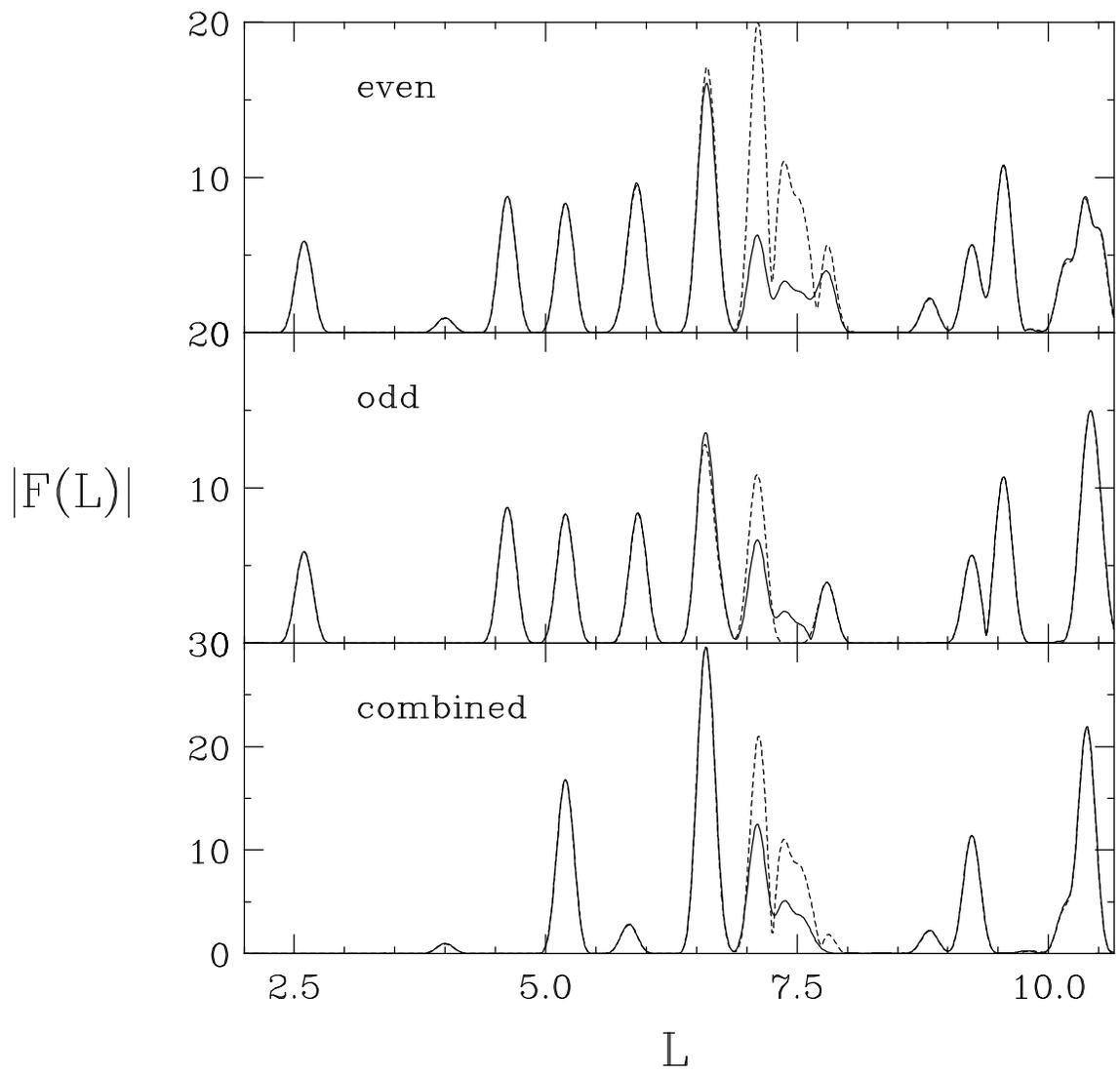
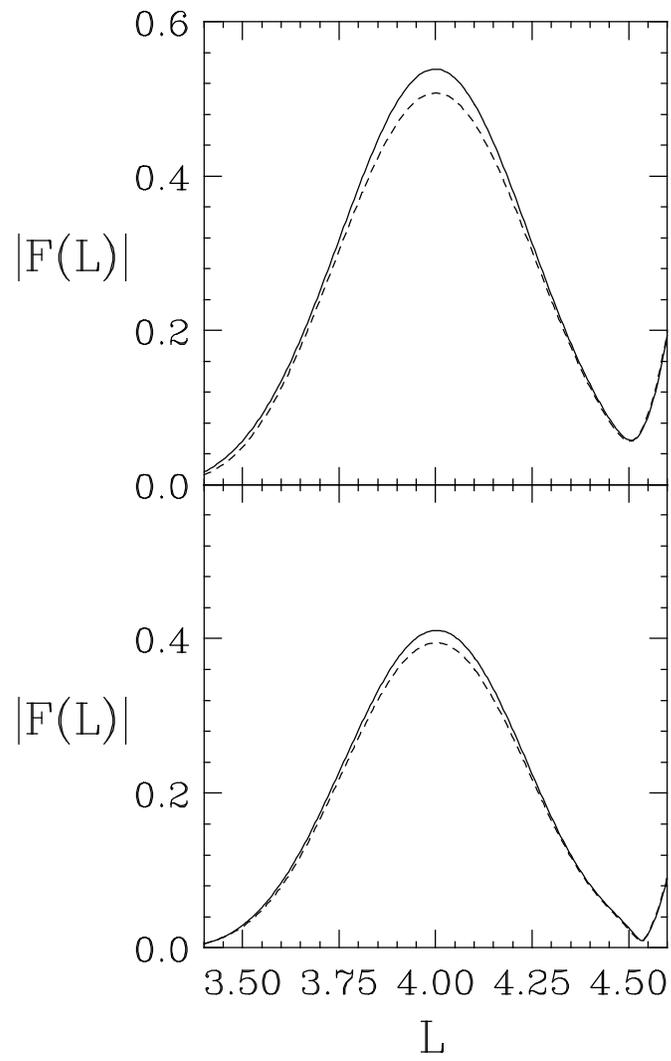
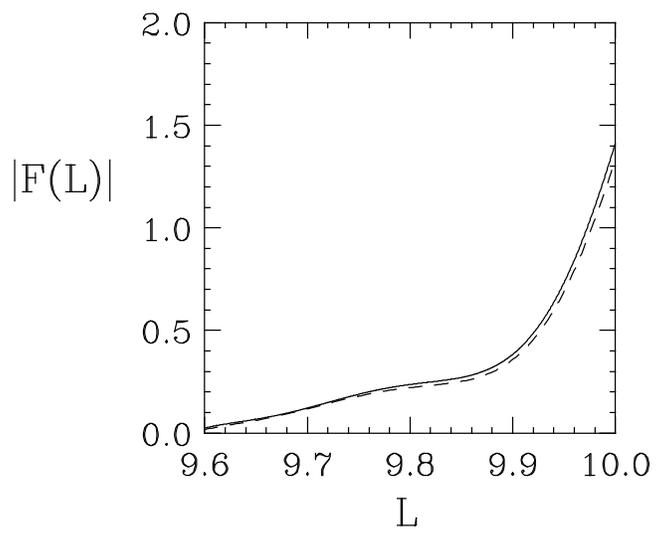
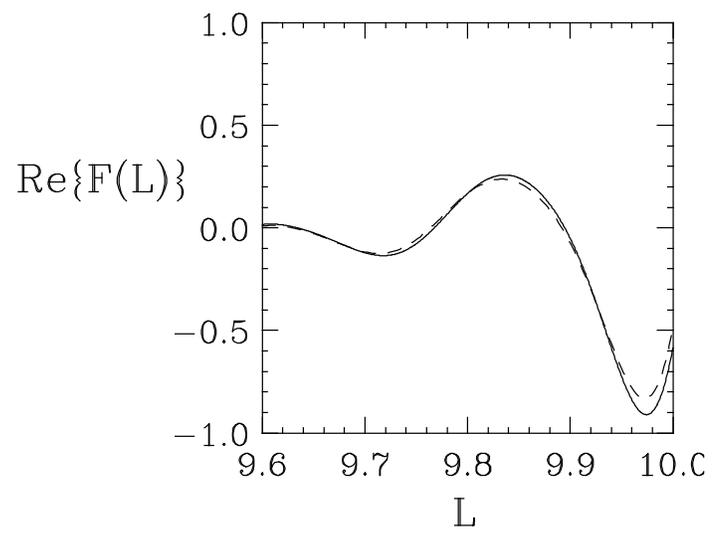
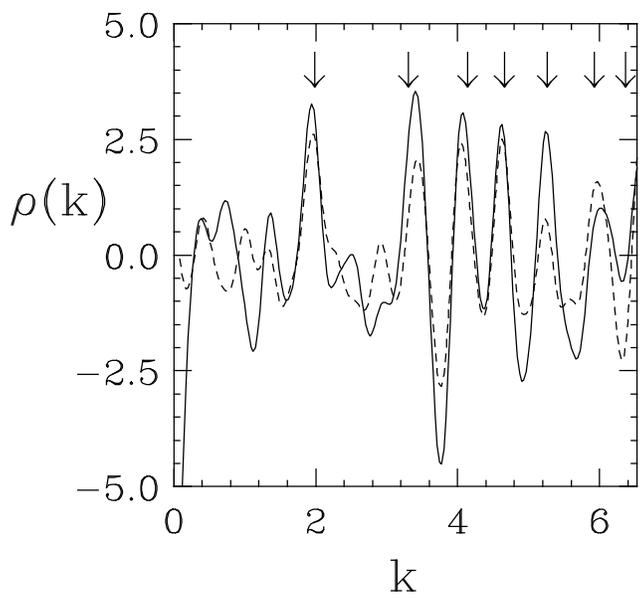
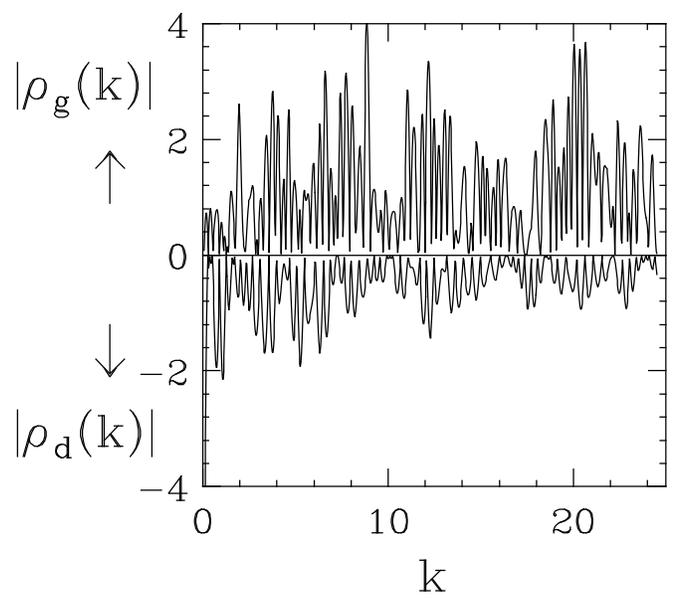
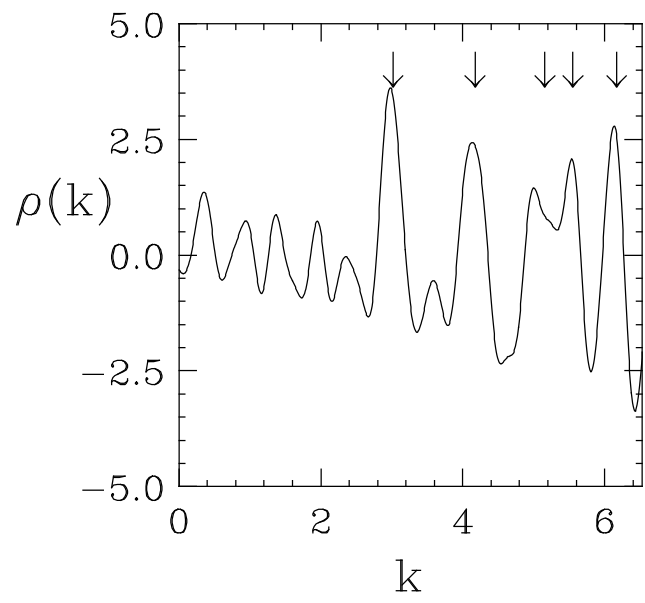
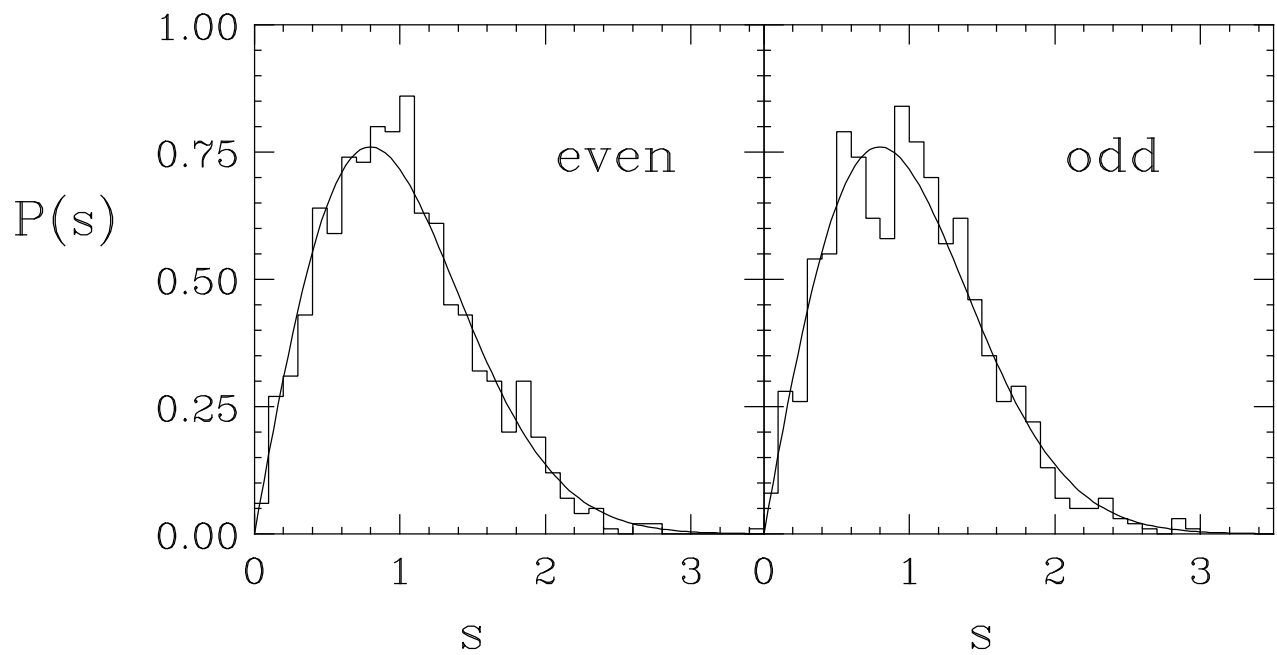
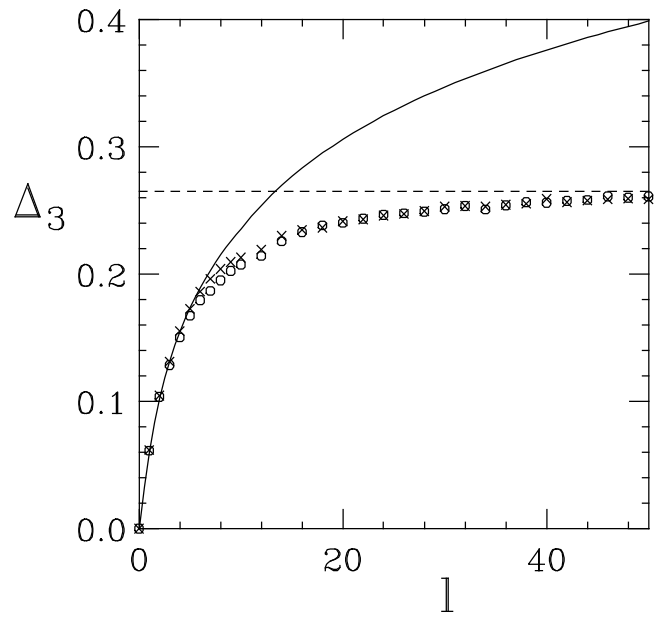
\end{document}